\documentclass[
  aps,
  prd,
  preprint,
  superscriptaddress,
  longbibliography
]{revtex4-2}

\usepackage[utf8]{inputenc}
\usepackage{graphicx}
\usepackage{amsmath,amssymb}
\usepackage{dsfont}
\usepackage{booktabs}

\usepackage[
  colorlinks=true,
  linkcolor=blue,
  citecolor=blue,
  urlcolor=blue
]{hyperref}

\def\FF{{\mathcal{F}}}
\def\HH{{\mathcal{H}}}
\def\B{{\mathcal{B}}}

\begin{document}

\title{Holographic Mutual Information in an Anisotropic Field Theory}
\author{Mehri Rahimi}
\email{m.rahimi.physics@gmail.com}
\affiliation{
School of Quantum Physics and Matter,
Institute for Research in Fundamental Sciences (IPM),
P.O. Box 19395-5531, Tehran, Iran
}

\author{Masoumeh Tavakoli}
\email{tavakoli.phy@gmail.com}
\affiliation{
Department of Physics,
Isfahan University of Technology,
84156-83111, Isfahan, Iran
}

\author{Behrouz Mirza}
\email{b.mirza@iut.ac.ir}
\affiliation{
Department of Physics,
Isfahan University of Technology,
84156-83111, Isfahan, Iran
}

\begin{abstract}
We investigate holographic mutual information in an anisotropic strongly coupled plasma within the framework of gauge/gravity duality. Considering both parallel and perpendicular strip configurations, we analyze the effects of temperature, anisotropy, strip lengths, and separation distance on the correlation structure between subsystems. For strips with equal lengths, we show that increasing temperature suppresses mutual information and reduces the size of the correlated region in the phase diagram. We find that the parallel configuration exhibits a monotonic behavior. However, the perpendicular configuration displays a more complex behavior resulting from the competition between the effects of energy density and pressure. Similar qualitative features are observed for strips with unequal lengths. Our results suggest that the directional dependence of the energy density and pressure plays an important role in the behavior of mutual information and may explain the nontrivial directional effects observed in the phase structure.
\end{abstract}

\maketitle

\section{Introduction}
The AdS/CFT correspondence was first proposed by Juan Maldacena in 1997~\cite{Maldacena}.
However, the foundation of gauge/gravity duality was laid even earlier through the work of other theorists exploring concepts within string theory and black hole physics~\cite{Bekenstein,Polchinski}.

Additionally, Gerard 't Hooft's work in 1973 on the large \(N_c\) limit of SU(\(N_c\)) gauge theories played a significant role. He identified a specific limit where \(N_c\) is taken large while keeping the quantity \(g^2_{YM}N_c\) fixed. This approach, known as the \emph{planar diagram theory}, provided valuable insights into the behavior of gauge theories in the large \(N_c\) limit and provided a foundation for the later development of gauge/gravity duality. 't Hooft's important contributions to quantum field theory and the renormalization of gauge theories were crucial for the theoretical framework on which Maldacena's AdS/CFT correspondence was built~\cite{tHooft1974}.

Maldacena proposed a duality between a conformal field theory (CFT) on the boundary of Anti-de Sitter (AdS) space and a string theory formulated in the bulk.
Over the past few decades, the gauge/gravity duality has evolved into a powerful theoretical framework, offering new insights into strongly coupled systems, quantum field theories, and the nature of space-time itself~\cite{Nishioka2009}. The duality has found applications in  various areas of physics, from condensed matter theory to cosmology, and continues to inspire ongoing research and discoveries.

In particle physics, gauge/gravity duality is instrumental in solving complex problems related to strongly coupled systems and understanding the behavior of quark--gluon plasma~\cite{CasalderreySolana2014}. This duality provides a powerful framework for studying the non-perturbative regime of gauge theories.
One of the most significant ongoing experimental programs is the study of heavy-ion collisions at colliders such as the Large Hadron Collider (LHC) and the Relativistic Heavy Ion Collider (RHIC). These collisions produce a quark--gluon plasma whose properties provide valuable insights into strongly interacting matter~\cite{CasalderreySolana2014}.

Our research is motivated by the fact that the quark--gluon plasma produced in heavy-ion collisions is both strongly coupled and anisotropic during its early evolution~\cite{CasalderreySolana2014,Mateos2011}. Standard methods, such as perturbation theory, are not applicable to strongly coupled theories. Consequently, gauge/gravity duality offers a promising approach for understanding the intricate properties of quark--gluon plasma and providing insight into the non-perturbative dynamics of quark--gluon plasma.

In addition to these considerations, entanglement entropy and mutual information are fundamental concepts in quantum information theory and play an important role in understanding quantum correlations and the structure of quantum states. Following the pioneering holographic proposals of Ryu and Takayanagi~\cite{Ryu2006a,Ryu2006b}, entanglement entropy has become one of the most powerful probes of strongly coupled quantum field theories. Comprehensive reviews of holographic entanglement entropy and its applications can be found in Refs.~\cite{Nishioka2009,Nishioka2018,Headrick2018}. It has been widely applied to both conformal and non-conformal holographic backgrounds, providing valuable insights into quantum correlations, phase transitions, and critical phenomena~\cite{Rahimi2017,Taylor2018,Mirza2019RT,Mirza2019Accelerating,Ebrahim2023}.
An important development in recent years has been the emergence of entanglement thermodynamics, which establishes a close connection between quantum entanglement and thermodynamic quantities. In particular, for sufficiently small subsystems, variations of the entanglement entropy satisfy a first-law-like relation involving the energy and other thermodynamic quantities of the system~\cite{Alishahiha2013,Mirza2019FirstLaw}.

More recent studies have further explored the connection between entanglement measures and thermal or chemical response functions, demonstrating that quantum entanglement can serve as a sensitive probe of the thermodynamic properties of strongly coupled systems~\cite{Jokela2026,Jokela2026Thermo}. These developments provide additional motivation for employing entanglement-based observables to study anisotropic strongly coupled plasmas.
Mutual information, on the other hand, quantifies the total correlations, both classical and quantum, between two subsystems. Unlike entanglement entropy, it is finite without requiring additional regularization and therefore provides a robust probe of quantum correlations in holographic systems. It has been employed to investigate the effects of energy scale and the number of degrees of freedom~\cite{Fischler2013,Rahimi2021,Oliveira2025}. In the context of holographic duality and anisotropic field theories, mutual information provides a valuable tool for probing the entanglement structure and the interplay between geometry and quantum information.

Our previous studies primarily focused on entanglement entropy in anisotropic systems~\cite{Rahimi2018}. Holographic mutual information has also been studied in several anisotropic backgrounds. Jahnke~\cite{Jahnke2018} computed the mutual information between strip-like regions on the two asymptotic boundaries of the maximally extended Mateos--Trancanelli black brane and examined how shock waves disrupt these correlations. Liu et al.~\cite{Liu2019Anisotropy} investigated the orientation dependence of mutual information in an anisotropic Q-lattice background. In this work, we consider two disjoint strips on the same boundary of the finite-temperature Mateos--Trancanelli geometry. For both parallel and perpendicular configurations, we determine the critical separation beyond which the mutual information vanishes and examine its dependence on temperature, anisotropy, and the strip lengths. We interpret the different responses of the two orientations in terms of the interplay between the energy density and the direction-dependent pressure.

\section{Theoretical Background}\label{sectionreviwe}
\subsection{Anisotropic Background Geometry}
An anisotropic background solving the type IIB supergravity equations
of motion was introduced in Ref.~\cite{Mateos2011}. The corresponding
five-dimensional metric in the Einstein frame is
\begin{align}
ds_E^2
&=
\frac{e^{-\phi(u)/2}}{u^2}
\left[
-\mathcal{F}(u)\mathcal{B}(u)\,dt^2
+dx^2+dy^2
+\mathcal{H}(u)\,dz^2
+\frac{du^2}{\mathcal{F}(u)}
\right],
\label{one}\\
\chi&=az,
\qquad
\phi=\phi(u).
\nonumber
\end{align}
Here, $\chi$ and $\phi$ are the axion and the dilaton fields, respectively, while $a$ denotes the anisotropy parameter, with dimensions of energy, related to the D7-brane density $n_{\mathrm{D}7}$ through

\begin{equation}
a=\frac{\lambda n_{\mathrm{D} 7}}{4 \pi N_{\mathrm{c}}}
\end{equation}
where $N_c$ is the number of colors of the gauge theory, and $\lambda=g_{\mathrm{YM}}^2N_c$ is the 't Hooft coupling of the gauge theory.
The functions $\HH$, $\FF$ and $\B$ depend only on the radial direction $u$. Throughout this work, the $AdS_5$ radius is set to unity. In terms of the dilaton field, they are given by
\begin{align}\label{three}
\HH&=e^{-\phi},\\
\label{ff} \FF&= \frac{e^{-\frac{1}{2}\phi}\left[ a^2 e^{\frac{7}{2}\phi}(4u+u^2\phi')+16\phi'\right]}{4(\phi'+u\phi'')}\,,
\\
\frac{\B'}{\B}&=\frac{1}{24+10
u\phi'}\left(24\phi'-9u\phi'^2+20u\phi''\right)\ .
\end{align}
 We take the horizon to be located at $u=u_h$ where $\FF (u_h)=0$ and the Hawking temperature
 % are 
 is given %by
 as follows:
\begin{equation}
T=-\frac{1}{4 \pi}\FF '(u_h)\sqrt{\B (u_h)}
\end{equation}
The dilaton field also satisfies the following third-order equation:
\begin{equation}
\label{eomphi}
\begin{aligned}
&\frac{256\phi'\phi''-16(\phi')^3(7u\phi'+32)}
       {u\,a^2 e^{7\phi/2}(u\phi'+4)+16\phi'}
+\frac{\phi'}{u(5u\phi'+12)(u\phi''+\phi')} \\
&\quad \times \Bigl[
13u^3(\phi')^4+8u\bigl(11u^2(\phi'')^2-60\phi''-12u\phi'''\bigr) \\
&\qquad
+u^2(\phi')^3\bigl(13u^2\phi''+96\bigr)
+2u(\phi')^2\bigl(-5u^3\phi'''+53u^2\phi''+36\bigr) \\
&\qquad
+\phi'\bigl(30u^4(\phi'')^2-64u^3\phi'''-288
             +32u^2\phi''\bigr)\Bigr]=0\,.
\end{aligned}
\end{equation}
Equation~(\ref{one}) is anisotropic in the $z$ direction unless $\HH=1$, while it remains isotropic in the $xy$ plane. In the dual field theory corresponding to this solution, the pressures in the $xy$ and $z$ directions are different.
\begin{equation}
\left\langle T_{i j}\right\rangle=\operatorname{diag}\left(E, P_{x y}, P_{x y}, P_z\right)
\end{equation}

\begin{equation}
P_{x y} \equiv P_x=P_y
\end{equation}

\begin{equation}
\begin{aligned}
E & =\frac{N_{\mathrm{c}}^2}{2 \pi^2}\left(-\frac{3}{4} \mathcal{F}_4-\frac{23}{28} \mathcal{B}_4+\frac{2777}{16128} a^4+\frac{c_{\mathrm{sch}}}{96} a^4\right), \\
P_{x y} & =\frac{N_{\mathrm{c}}^2}{2 \pi^2}\left(-\frac{1}{4} \mathcal{F}_4-\frac{5}{28} \mathcal{B}_4+\frac{611}{16128} a^4-\frac{c_{\mathrm{sch}}}{96} a^4\right), \\
P_z & =\frac{N_{\mathrm{c}}^2}{2 \pi^2}\left(-\frac{1}{4} \mathcal{F}_4-\frac{13}{28} \mathcal{B}_4+\frac{2227}{16128} a^4+\frac{c_{\mathrm{sch}}}{32} a^4\right) .
\end{aligned}
\end{equation}
Further details and definitions of $\mathcal{B}_4$, $\mathcal{F}_4$ and $c_{\mathrm{sch}}$ can be found in~\cite{Mateos2011}.
  %%can be obtained from the holographic stress tensor,

\subsection{Entanglement Entropy}\label{EE}
Entanglement entropy is a fundamental quantity in quantum information theory and quantum field theory. It quantifies the quantum correlations between different parts of a quantum system. Consider a quantum system described by a pure state $|\psi\rangle$, divided into two subsystems $A$ and $B$. The total Hilbert space of the system is given by:
\[
\mathcal{H} = \mathcal{H}_A \otimes \mathcal{H}_B.
\] The density matrix of the entire system is given by
\[
\rho = |\psi\rangle \langle \psi|.
\]
The reduced density matrix for subsystem $A$ is obtained by tracing over the degrees of freedom of subsystem $B$:
\[
\rho_A = \mathrm{Tr}_B (\rho).
\]

The entanglement entropy of subsystem $A$ is then defined as the von Neumann entropy of the reduced density matrix:
\begin{equation}
S_A = - \mathrm{Tr} (\rho_A \log \rho_A).
\end{equation}

In the gauge/gravity duality framework, the holographic entanglement entropy is computed via the Ryu–Takayanagi prescription~\cite{Ryu2006a,Ryu2006b}

\begin{equation}
\label{rt}
S_A = \frac{\text{Area}(\gamma_A)}{4 G_N^{(d+2)}},
\end{equation}  

\begin{figure}[!ht]
\centering
\includegraphics[width=70mm]{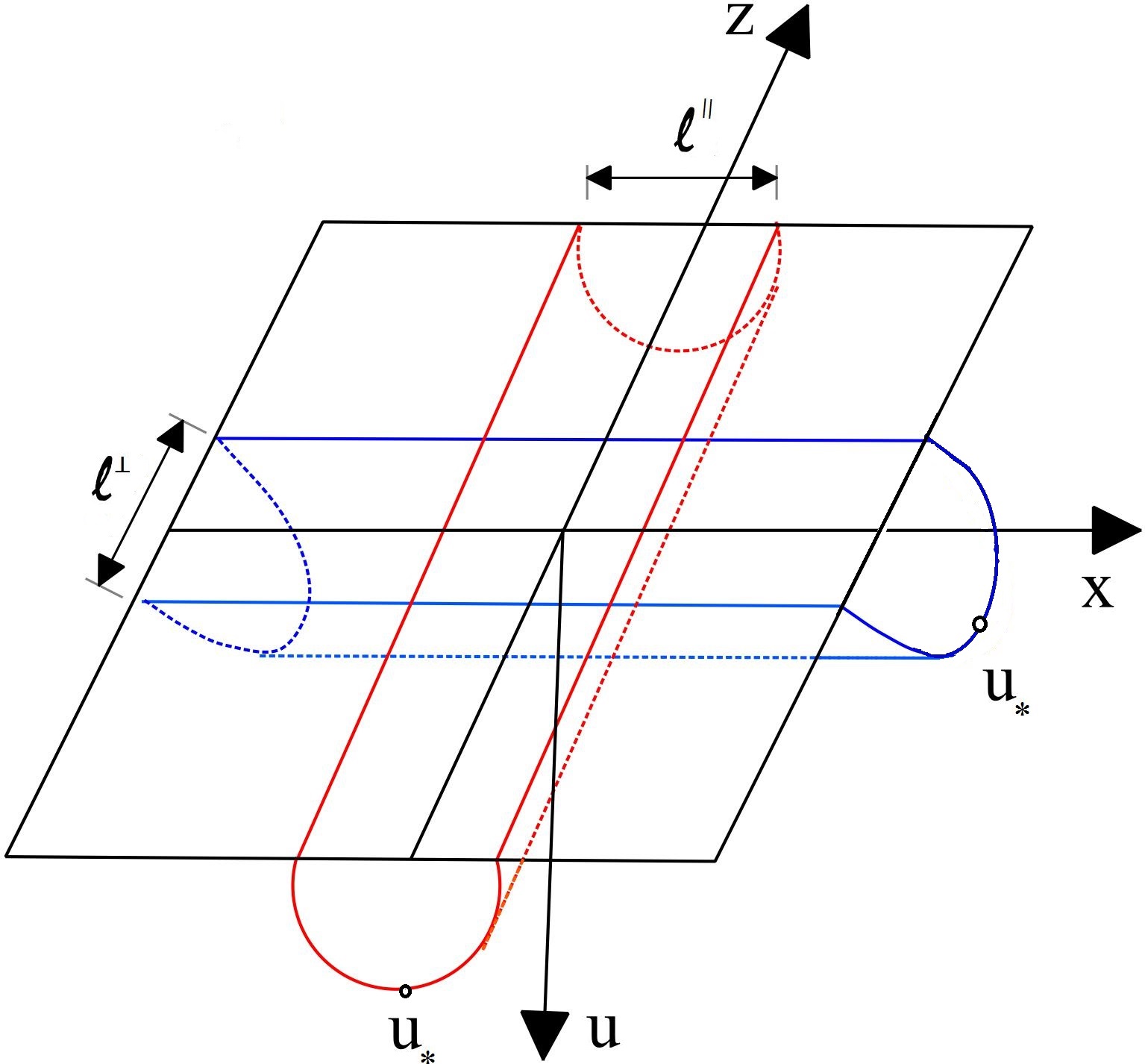}
\caption{Minimal bulk surfaces for the strips configuration in the anisotropic geometry.}
\label{configuration}
\end{figure}

where $G_N^{(d+2)}$ is the $(d+2)$-dimensional Newton constant and
$\gamma_A$ is the codimension-two extremal surface anchored on
$\partial A$. Throughout this work, we set $d=3$. We consider a
rectangular strip with finite length $l$ and regulated extent $L$ along each of the two remaining spatial directions, with $L\gg l$. The corresponding regulated transverse area is therefore $V_2=L^2$. Because of the anisotropy, two distinct orientations of the strip, as shown in Fig.~\ref{configuration}, must be considered.
\begin{itemize}
\item
\textbf{Parallel configuration:}
The strip has finite length $l$ along the anisotropic $z$-direction,
$-l/2<z<l/2$, and regulated extent $L$ along both the $x$ and $y$
directions. The entanglement entropy and strip length are
\begin{align}
S_A^{\parallel}
&=
\frac{V_2}{2G_N^{(5)}}
\int_{\epsilon}^{u_*}
\frac{e^{-3\phi(u)/2}\sqrt{\mathcal{H}(u)}}{\sqrt{\mathcal{F}(u)}\,u^6\sqrt{\frac{e^{-3\phi(u)/2}\mathcal{H}(u)}{u^6}
-
\frac{e^{-3\phi_*/2}\mathcal{H}_*}{u_*^6}}}
\,du,
\label{ep}\\
l^{\parallel}
&=
2\int_0^{u_*}
\frac{du}{\sqrt{\mathcal{F}(u)\mathcal{H}(u)}\sqrt{\frac{e^{-3\phi(u)/2}\mathcal{H}(u)u_*^6}{e^{-3\phi_*/2}\mathcal{H}_*u^6}-1}}.
\label{lp}
\end{align}
\item
\textbf{Perpendicular configuration:}
The strip has finite length $l$ along the $x$-direction,
$-l/2<x<l/2$, and regulated extent $L$ along the $y$ and $z$
directions. The entanglement entropy and strip length are
\begin{align}
S_A^{\perp}
&=
\frac{V_2}{2G_N^{(5)}}
\int_{\epsilon}^{u_*}\frac{e^{-3\phi(u)/2}\mathcal{H}(u)}{\sqrt{\mathcal{F}(u)}\,u^6\sqrt{\frac{e^{-3\phi(u)/2}\mathcal{H}(u)}{u^6}
-
\frac{e^{-3\phi_*/2}\mathcal{H}_*}{u_*^6}}}\,du,
\label{ebot}\\
l^{\perp}
&=
2\int_0^{u_*}
\frac{du}{\sqrt{\mathcal{F}(u)}
\sqrt{\frac{e^{-3\phi(u)/2}\mathcal{H}(u)u_*^6}{
e^{-3\phi_*/2}\mathcal{H}_*u^6}-1}}.
\label{lbot}
\end{align}
\end{itemize}
We denote the turning point of the extremal surface by $u_*$ and use
\[
\phi_*=\phi(u_*),
\qquad
\mathcal{H}_*=\mathcal{H}(u_*).
\]
A UV cutoff $u=\epsilon$ is introduced to regulate the divergent
entanglement entropy.

\subsection{Mutual Information}\label{MIsection}

Mutual information quantifies the total correlations between two
disjoint subsystems $A$ and $B$ and is defined as
\begin{equation}
I(A,B)=S_A+S_B-S_{A\cup B},
\label{MI}
\end{equation}
where $S_A$, $S_B$, and $S_{A\cup B}$ denote the entanglement
entropies of $A$, $B$, and their union, respectively.

\begin{figure}[!ht]
\centering
\includegraphics[width=140mm]{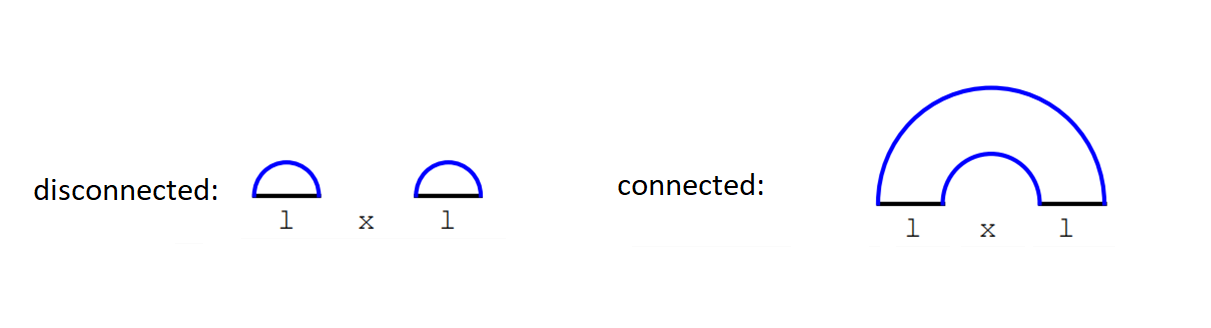}
\caption{Two candidate minimal surface configurations contributing to $S_{A \cup B}$. The phase transition in mutual information is associated with switching between the connected and disconnected configurations.}
\label{2configurations}
\end{figure}

Although the subsystems may have arbitrary shapes, in this work we
consider two rectangular strips of lengths $l_1$ and $l_2$, separated
by a distance $x$. As illustrated in Fig.~\ref{2configurations}, two
candidate minimal-surface configurations contribute to
$S_{A\cup B}$: a disconnected configuration and a connected one.
Their corresponding entanglement entropies are
$S(l_1)+S(l_2)$ and $S(x)+S(l_1+l_2+x)$, respectively. Therefore,
the entropy of the union is determined by
\begin{equation}
S_{A\cup B}
=
\min\left\{
S(l_1)+S(l_2),\,
S(x)+S(l_1+l_2+x)
\right\}.
\label{SAunionB_general}
\end{equation}

A phase transition occurs when the two configurations have equal
entanglement entropy. Accordingly, the critical separation
$x_{\mathrm{crit}}(l_1,l_2)$ is determined by
\begin{equation}
S(l_1)+S(l_2)
=
S\bigl(x_{\mathrm{crit}}\bigr)
+
S\bigl(l_1+l_2+x_{\mathrm{crit}}\bigr),
\label{xcrit_general}
\end{equation}
where, for brevity, the dependence of $x_{\mathrm{crit}}$ on
$l_1$ and $l_2$ is implicit on the right-hand side.

Using Eqs.~\eqref{MI} and \eqref{SAunionB_general}, the mutual
information can be written as
\begin{equation}
I(A,B)
=
\max\left\{
0,\,
S(l_1)+S(l_2)-S(x)-S(l_1+l_2+x)
\right\}.
\label{MI_general}
\end{equation}
Equivalently,
\begin{equation}
I(A,B)=
\begin{cases}
0,
& x\geq x_{\mathrm{crit}}(l_1,l_2),\\[2mm]
S(l_1)+S(l_2)-S(x)-S(l_1+l_2+x),
& x<x_{\mathrm{crit}}(l_1,l_2).
\end{cases}
\label{MI_general_piecewise}
\end{equation}

For two strips of equal length, $l_1=l_2=l$, the criticality
condition reduces to
\begin{equation}
2S(l)
=
S\bigl(x_{\mathrm{crit}}(l)\bigr)
+
S\bigl(2l+x_{\mathrm{crit}}(l)\bigr),
\label{MI0}
\end{equation}
and the mutual information becomes
\begin{equation}
I(A,B)=
\max\left\{
0,\,
2S(l)-S(x)-S(2l+x)
\right\}.
\label{MI_equal}
\end{equation}

All the above expressions apply separately to the parallel and
perpendicular configurations by using the corresponding entanglement
entropy, $S^{\parallel}$ or $S^{\perp}$.

\section{Numerical Results}\label{numeric}

In this section, we investigate the effects of various parameters, such as the anisotropy parameter, temperature, the strip length $l$, and the separation distance $x$, on the mutual information and the critical separation $x_{\text{crit}}$. The regions $A$ and $B$ are taken to be rectangular strips, with either equal or unequal lengths. We analyze both cases in Sections~\ref{equal} and \ref{unequal}, respectively. In all plots presented in this section, black curves represent the isotropic conformal case, while colored curves correspond to the anisotropic background with different values of $a$ and $T$.

\subsection{Strips with Equal Lengths}\label{equal}

In this subsection, we study the mutual information between two parallel strips with equal lengths, $l_1=l_2=l$, separated by a distance $x$, in the anisotropic background. We investigate the effects of temperature\footnote{Strictly speaking, the relevant physical quantity is the dimensionless ratio $\dfrac{T}{\mu}$, where $\mu$ is an energy scale. Throughout this paper, we set $\mu = 1$, so the temperature dependence can be discussed directly in terms of $T$.} and anisotropy on the mutual information and the associated phase transition, and compare the results with those of the isotropic conformal field theory. We first focus on temperature effects and then discuss the role of the anisotropy parameter.

\subsubsection{Effect of Temperature}
\begin{figure}[!ht]
\begin{center}
\includegraphics[width=70mm]{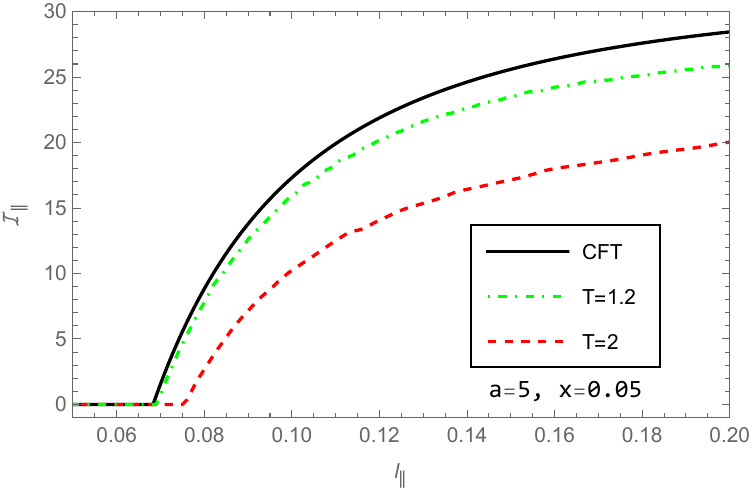}
	\hspace{0.5cm}
\includegraphics[width=70mm]{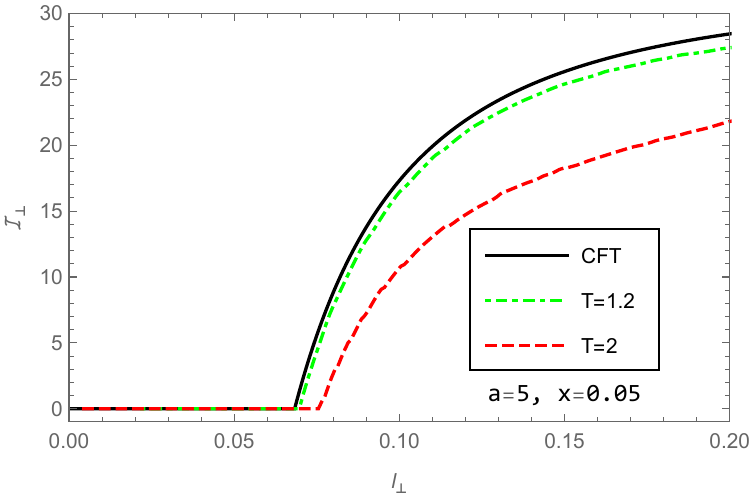}
%Iasafunctionoflfixedaaniso}
\caption{Mutual information as a function of the strip length $l$ for $a=5$ at $T=1.2$ and $T=2$, with $x=0.05$, compared with the isotropic conformal case.}
\label{I(l)fixeda}
\end{center}
\end{figure}

\begin{figure}[!ht]
\begin{center}
\includegraphics[width=70mm]{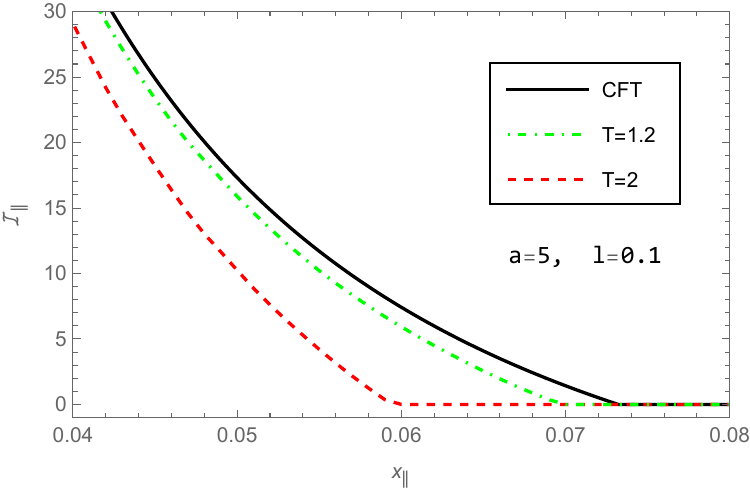}
	\hspace{0.5cm}
\includegraphics[width=70mm]{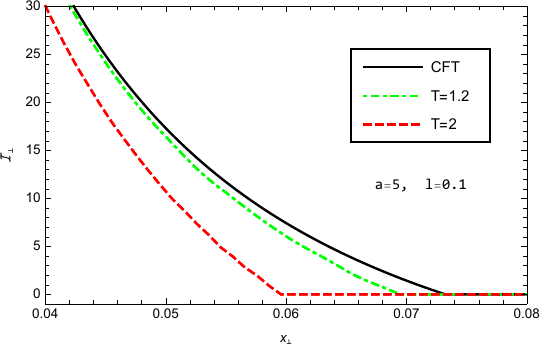}
%Iasafunctionofxfixedaaniso}
\caption{Mutual information as a function of the separation distance $x$ for $a=5$, with $l=0.1$ at $T=1.2$ and $T=2$, compared with the isotropic conformal case.}
\label{I(x)fixeda}
\end{center}
\end{figure}

We begin by analyzing the temperature dependence of the mutual information for the symmetric strip configuration. Figs.~\ref{I(l)fixeda} and \ref{I(x)fixeda} show the mutual information as functions of the strip length $l$ and the separation distance $x$, respectively, for a fixed anisotropy parameter $a=5$. The results are compared with the zero-temperature isotropic conformal background. The behavior in the low-temperature regime is discussed separately in Appendix~B.

\begin{figure}[!ht]
\begin{center}
\includegraphics[width=70mm]{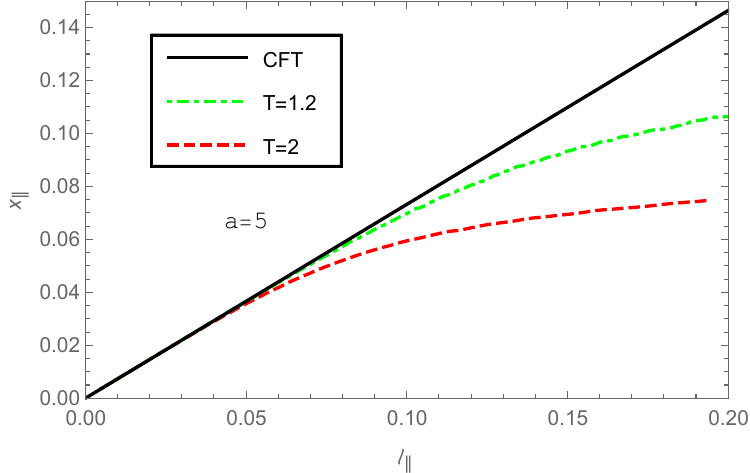}
	\hspace{0.5cm}
\includegraphics[width=70mm]{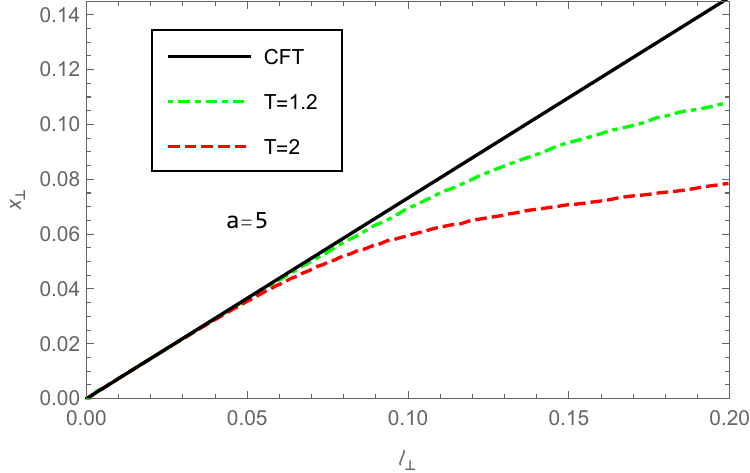}
\caption{Phase diagram of mutual information for anisotropy parameter $a=5$ at temperatures $T=1.2$ and $T=2$, compared with the isotropic conformal case. The curves separate the regions of non-zero and vanishing mutual information.}\label{phasea5}
\end{center}
\end{figure}

In Fig.~\ref{I(l)fixeda}, the mutual information is plotted for different temperatures, namely $T=1.2$, $T=2$, and the isotropic zero-temperature conformal background. For the temperatures considered here, the mutual information increases with increasing strip length, as larger subsystems probe more degrees of freedom and thus capture stronger correlations. However, the finite-temperature anisotropic states exhibit smaller mutual information compared to the zero-temperature  isotropic conformal case. This behavior reflects the suppression of quantum correlations by thermal fluctuations~\cite{Zurek1991}.

Similarly, Fig.~\ref{I(x)fixeda} shows that the mutual information decreases as the separation distance $x$ increases and eventually vanishes, indicating a disentangling phase transition. As discussed earlier, for each strip length $l$, there exists a corresponding critical separation $x_{\text{crit}}(l)$ beyond which the mutual information vanishes. Moreover, increasing the temperature reduces the critical separation associated with a given strip length, meaning that the connected phase becomes less favorable and the correlated region in the phase diagram becomes smaller. 

This behavior can be understood in terms of the increase in the energy density of the system at higher temperatures. Previous studies have shown that increasing the energy density suppresses the mutual information~\cite{Alishahiha2015}. Since the energy density increases with temperature, the mutual information between two static regions is maximal in the vacuum state and decreases as the temperature increases.

In Fig.~\ref{phasea5}, the phase diagrams for fixed $a=5$ and different temperatures are shown. The curves represent the combinations of $x$ and $l$ at which a first-order phase transition occurs. The region below each curve corresponds to configurations with non-zero mutual information, whereas the region above the curves corresponds to vanishing mutual information. As noted in~\cite{Fischler2013}, the parameter space of the correlated phase decreases with increasing temperature, which is reflected in the corresponding shift of the critical curve.

In other words, increasing temperature reduces the range of separations for which the two subsystems remain correlated. This can be understood from the competition between thermal fluctuations and quantum entanglement: increasing temperature enhances thermal effects and suppresses quantum correlations. From a physical perspective, this corresponds to a reduction in the correlation length of the system, leading to a faster decay of correlations between spatially separated regions.

We note that the role of anisotropy becomes more intricate when different orientations of the strips are considered, as will be discussed in the following subsection.
 \begin{figure}
	\begin{center}
	\includegraphics[width=70mm]{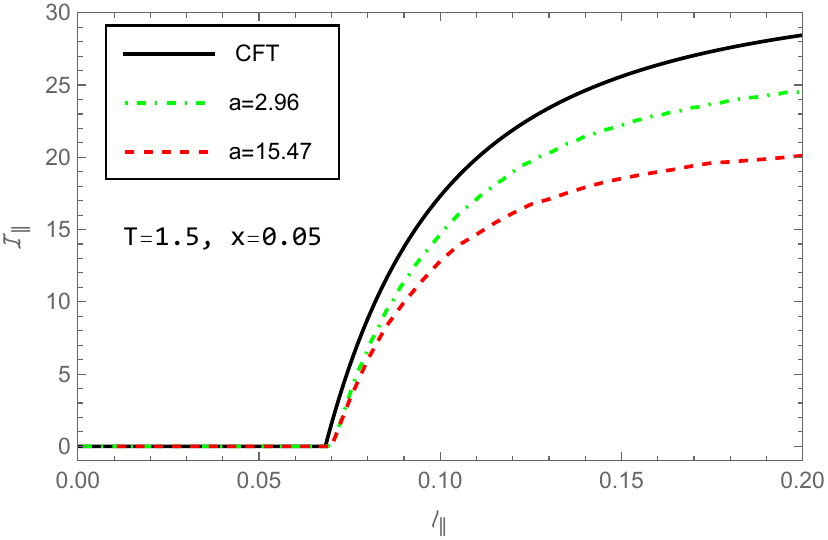}
	\hspace{0.5cm}
	\includegraphics[width=70mm]{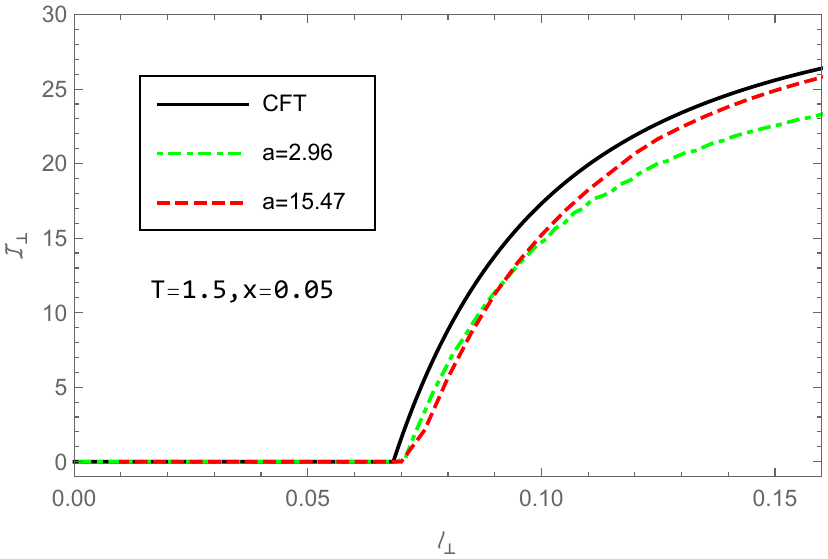}
    \caption{Mutual information as a function of the strip length $l$ for anisotropy parameters $a=2.96$ and $a=15.47$, at fixed
temperature $T=1.5$ and separation distance $x=0.05$, compared
with the isotropic conformal case.}\label{MIlTfixed}
	\end{center}
\end{figure}

\begin{figure}[!ht]
	\begin{center}
		\includegraphics[width=70mm]{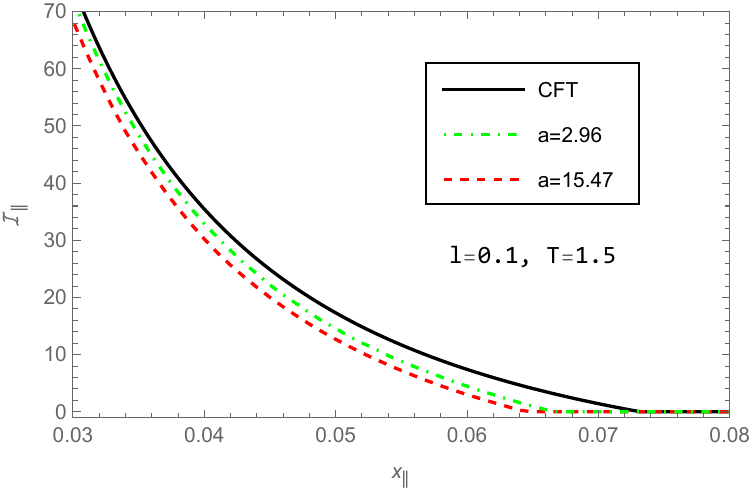}
					\hspace{0.5cm}
		\includegraphics[width=70mm]{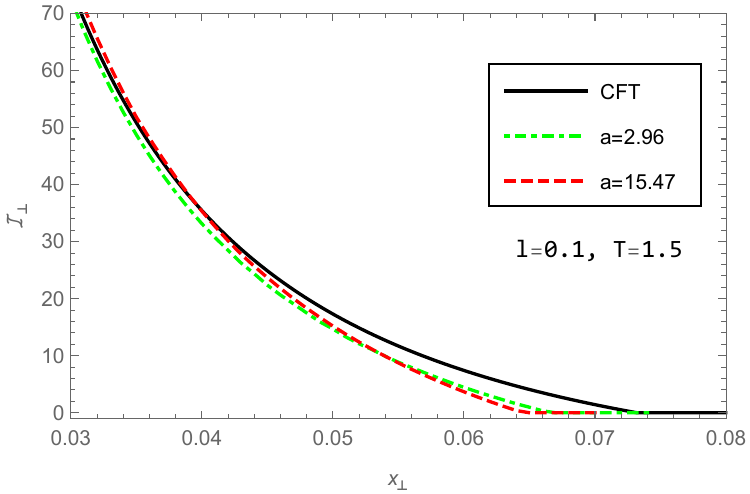}
		\caption{Mutual information as a function of the separation distance
$x$ for anisotropy parameters $a=2.96$ and $a=15.47$, at fixed
temperature $T=1.5$ and strip length $l=0.1$, compared with the
isotropic conformal case.}\label{MIxTfixed}
	\end{center}
\end{figure}

\begin{figure}[h]
\begin{center}
\includegraphics[width=70mm]{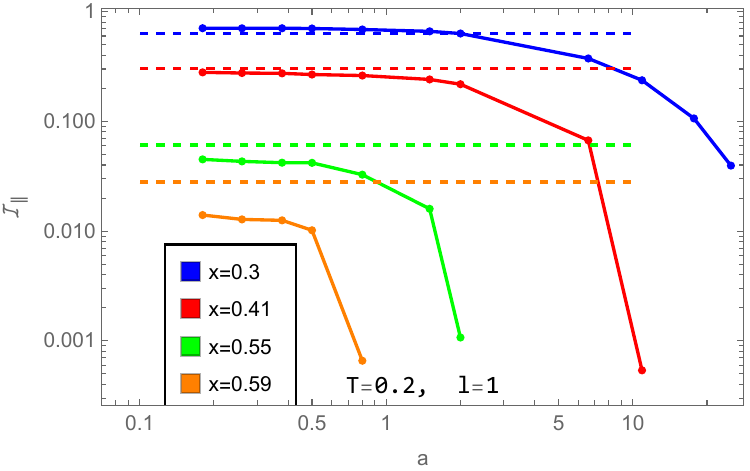}
\hspace{0.5cm}
\includegraphics[width=70mm]{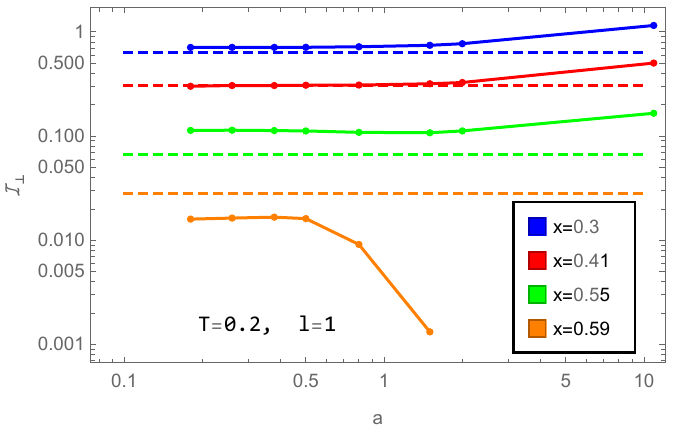}
\caption{Mutual information as a function of the anisotropy parameter for different values of the separation distance $x$, at fixed temperature $T=0.2$ and strip length $l=1$, compared with the isotropic conformal case at the same temperature.}\label{T02I(a)}
\end{center}
\end{figure}

\begin{figure}[h]
\begin{center}
\includegraphics[width=70mm]{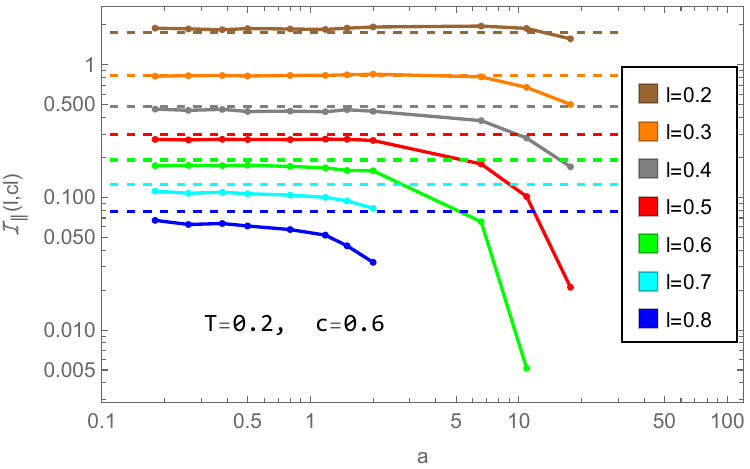}
\hspace{0.5cm}
\includegraphics[width=70mm]{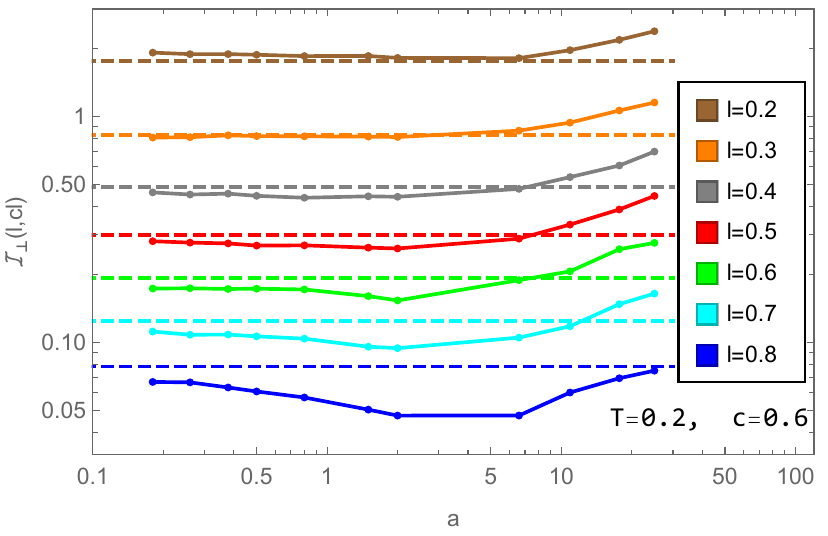}
\caption{Mutual information $I(l,cl)$ with $c=0.6$ as a function of the anisotropy parameter at $T=0.2$, for different values of the strip length $l$. The horizontal dashed lines correspond to the isotropic conformal case under the same conditions.}\label{Icl2}
\end{center}
\end{figure}

\subsubsection{Effect of the Anisotropy Parameter}
In this subsection, we study the effect of the anisotropy parameter on the mutual information and the phase structure of the system at fixed temperature. Our goal is to understand how increasing anisotropy modifies the correlations between the subsystems and gives rise to different behaviors in the parallel and perpendicular configurations.

We first analyze the mutual information as a function of the strip length $l$ and the separation distance $x$, shown in Figs.~\ref{MIlTfixed} and \ref{MIxTfixed}. In the parallel configuration, the mutual information exhibits a monotonic behavior with respect to the anisotropy parameter, indicating a consistent deviation from the isotropic conformal field theory. 

In contrast, the perpendicular configuration exhibits a more intricate behavior. For small values of the strip length, increasing anisotropy enhances the deviation from the isotropic conformal result, whereas for sufficiently large $l$, the deviation is reduced. A similar non-monotonic behavior is observed as a function of the separation distance: near the critical region (larger separations), increasing anisotropy leads to a stronger deviation, while for smaller separations the deviation decreases, eventually crossing the isotropic conformal curve and then increasing again. This indicates that the effect of anisotropy is not determined solely by the parameter $a$, but depends on additional physical quantities.

To understand these behaviors, it is useful to consider the roles of energy density and pressure. As shown in Ref.~\cite{Mateos2011}, increasing the anisotropy parameter always increases the energy density. However, the pressure depends on the direction: the longitudinal pressure (i.e., the pressure along the anisotropic direction) increases with anisotropy, whereas the transverse pressure (i.e., the pressure perpendicular to the anisotropic direction) decreases. Within the framework of entanglement thermodynamics, variations of the entanglement entropy are related to both the energy density and the pressure of the system~\cite{Alishahiha2013,Mirza2019FirstLaw}. Furthermore, previous studies have shown that increasing the energy density suppresses the mutual information~\cite{Alishahiha2015}. Motivated on these results, we interpret our numerical findings by taking into account the directional dependence of the pressure in the anisotropic background.

Our results are fully consistent with this general picture. The distinctive feature of the present anisotropic background is that increasing the anisotropy parameter affects the pressure differently in the two directions. In the parallel configuration, both the energy density and the longitudinal pressure increase with anisotropy, so their effects reinforce each other, leading to a monotonic decrease of the mutual information. In contrast, in the perpendicular configuration, the energy density increases whereas the transverse pressure decreases. Consequently, the effects of energy density and pressure act in opposite directions and compete, giving rise to the observed non-monotonic behavior.

To further illustrate this competition, we examine the dependence of the mutual information on the anisotropy parameter. Fig.~\ref{T02I(a)} shows the mutual information as a function of the anisotropy parameter for a fixed temperature $T=0.2$, strip length $l=1$, and several values of the separation distance $x$. For the parallel configuration, the mutual information decreases monotonically with increasing anisotropy for all values of $x$, consistent with the discussion above.

In the perpendicular configuration, however, the behavior depends sensitively on the separation distance. For larger separations, the mutual information decreases with increasing anisotropy, similar to the parallel case. In contrast, for smaller separations, it becomes larger than the isotropic conformal result and increases with the anisotropy parameter. This indicates that, in this regime, the reduction of the transverse pressure outweighs the suppressing effect of the increasing energy density, leading to enhanced correlations between the subsystems.

A natural question is whether the separation distance alone determines this behavior or whether the relevant quantity is the ratio $\frac{x}{l}$. To address this question, we study configurations with fixed values of this ratio, as shown in Fig.~\ref{Icl2}. In particular, we consider $x=cl$ with $c=0.6$ at $T=0.2$, where $c=0.6$ is close to the critical ratio of the isotropic conformal field theory.

\begin{figure}[h]
\begin{center}
\includegraphics[width=70mm]{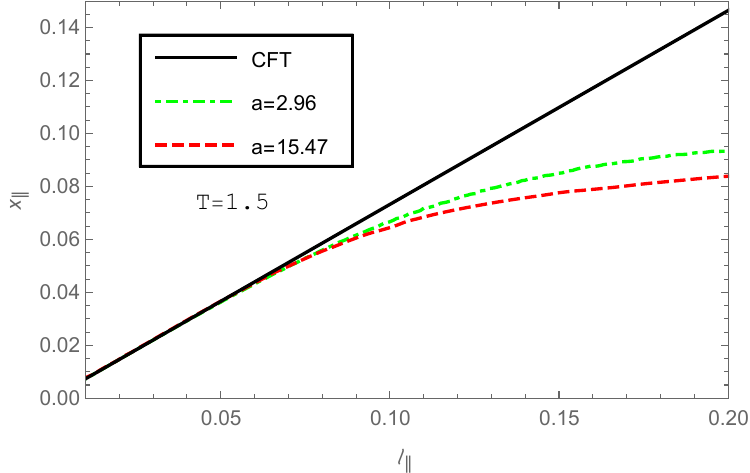}
\hspace{0.5cm}
\includegraphics[width=70mm]{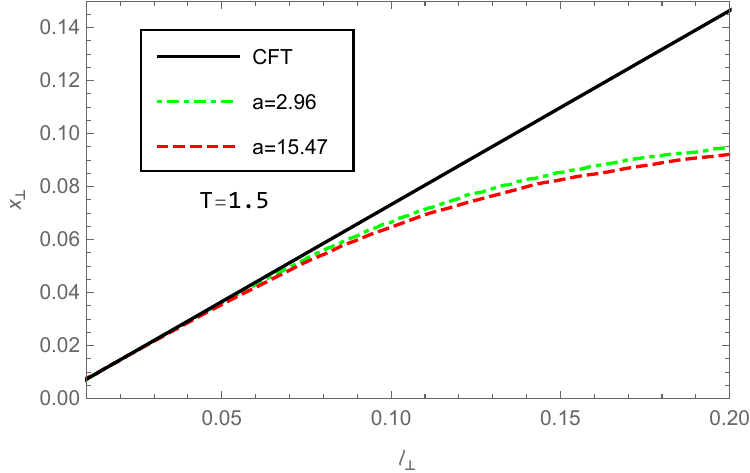}
\caption{Phase diagram for two strips with equal length $l$ and separation distance $x$ at temperature $T=1.5$, for anisotropy parameters $a=2.96$ and $a=15.47$, compared with the zero-temperature isotropic conformal case.} \label{phaseTfixeddiffa}
\end{center}
\end{figure}

We find that the mutual information depends not only on the ratio $\frac{x}{l}$ but also on the absolute values of $l$ and $x$, indicating that the ratio alone is insufficient to characterize the system. In the perpendicular configuration, for larger strip lengths $l$ (and correspondingly larger separations $x$), the mutual information initially decreases as the anisotropy increases and then begins to rise. In contrast, for smaller values of $l$, the mutual information exhibits an almost monotonic increase with the anisotropy parameter. These results are consistent with the competition between the increasing energy density and the decreasing transverse pressure discussed above, whose relative importance depends on the absolute values of $l$ and $x$, rather than only on their ratio.
Finally, we analyze the phase diagrams for different values of the anisotropy parameter at fixed temperature, shown in Fig.~\ref{phaseTfixeddiffa}. The curves represent the critical points in the $(x,l)$ plane where the transition between connected and disconnected entangling surfaces occurs. The region below each curve corresponds to non-zero mutual information, while the region above corresponds to vanishing mutual information.

We observe that increasing the anisotropy parameter reduces the size of the correlated region. Equivalently, for a given strip length $l$, the corresponding critical separation $x_{\text{crit}}(l)$ becomes smaller as the anisotropy increases. This behavior is consistent with the overall effect of increasing energy density, which suppresses correlations, together with the directional modification of pressure in the anisotropic background.

\subsection{Strips with Unequal Lengths}\label{unequal}
In this section, we consider two strips of unequal lengths, $l_1$ and
$l_2$, separated by a distance $x$. Their mutual information is computed
using Eq.~\eqref{MI_general}, while the corresponding phase boundary is
determined by Eq.~\eqref{xcrit_general}.
\subsubsection{Effect of Temperature}

\begin{figure}[h]
\begin{center}
\includegraphics[width=70mm]{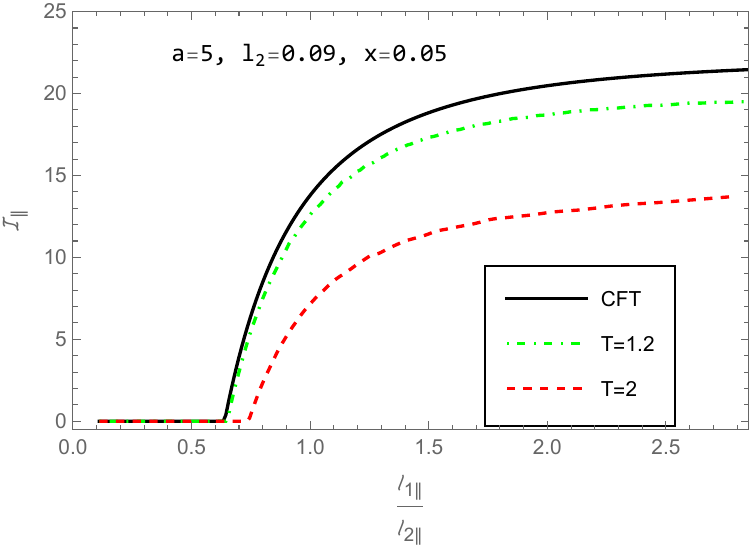}
\hspace{0.5cm}
\includegraphics[width=70mm]{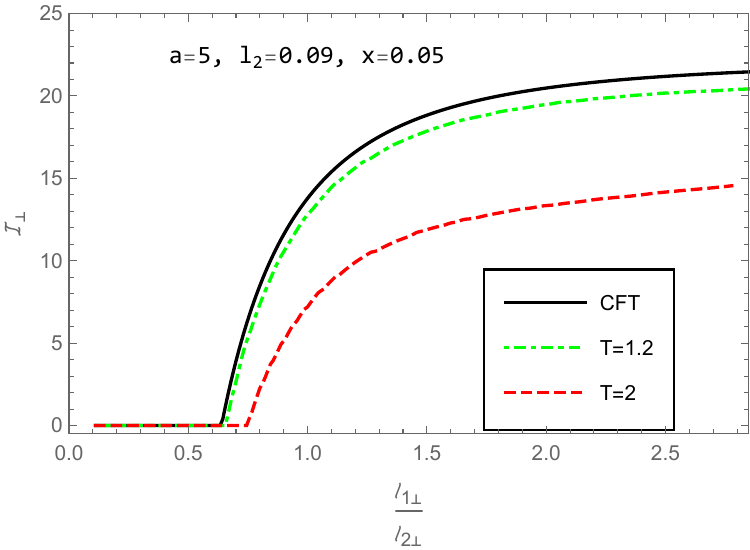}
\caption{Mutual information as a function of the ratio $\frac{l_1}{l_2}$ for different temperatures $T=1.2$ and $T=2$, at fixed anisotropy parameter $a=5$, strip length $l_2=0.09$, and separation distance $x=0.05$, compared with the isotropic conformal case.}  \label{MIunequalT}
\end{center}
\end{figure}

In this subsection, we study the effect of temperature on the mutual information for strips with unequal lengths $l_1$ and $l_2$. In Fig.~\ref{MIunequalT}, the mutual information is shown as a function of the ratio $\frac{l_1}{l_2}$ for different temperatures at fixed anisotropy parameter $a=5$, with $l_2=0.09$ and separation distance $x=0.05$, compared with the isotropic conformal case.

As expected, the mutual information decreases with increasing temperature, in agreement with the behavior observed for equal-length strips. This reflects the suppression of correlations by thermal fluctuations.

\begin{figure}[h]
\begin{center}
\includegraphics[width=70mm]{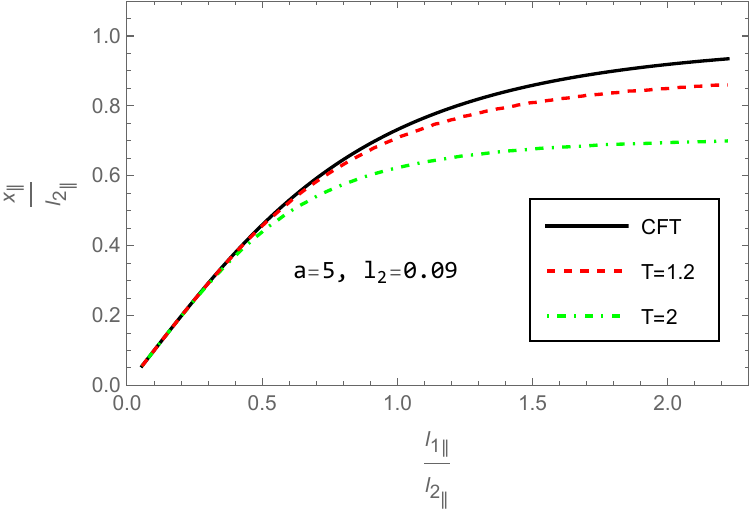}
\hspace{0.5cm}
\includegraphics[width=70mm]{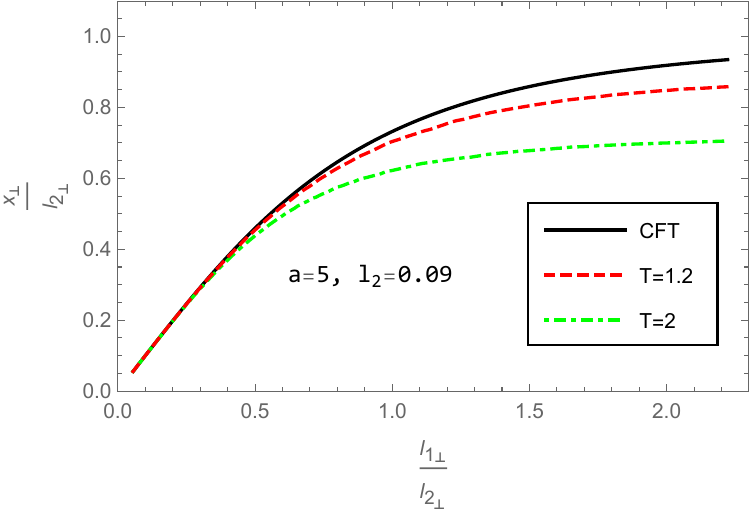}
\caption{Phase diagram for two strips with unequal lengths, plotted as $\frac{x}{l_2}$ versus $\frac{l_1}{l_2}$, for $a=5$ at $T=1.2$ and $T=2$. The length $l_2=0.09$ is fixed, and results are compared with the isotropic conformal case.}  \label{phaseunequaldiffT}
\end{center}
\end{figure}

The corresponding phase diagram is shown in Fig.~\ref{phaseunequaldiffT}. Similar to the equal-length case, increasing the temperature reduces the size of the correlated region. The phase boundaries move toward smaller values of $\frac{x}{l_2}$, indicating that the connected phase becomes less favorable. Consequently, the parameter space in which mutual information remains non-zero becomes smaller as the temperature increases.

\subsubsection{Effect of the Anisotropy Parameter}
\begin{figure}[h]
\begin{center}
\includegraphics[width=70mm]{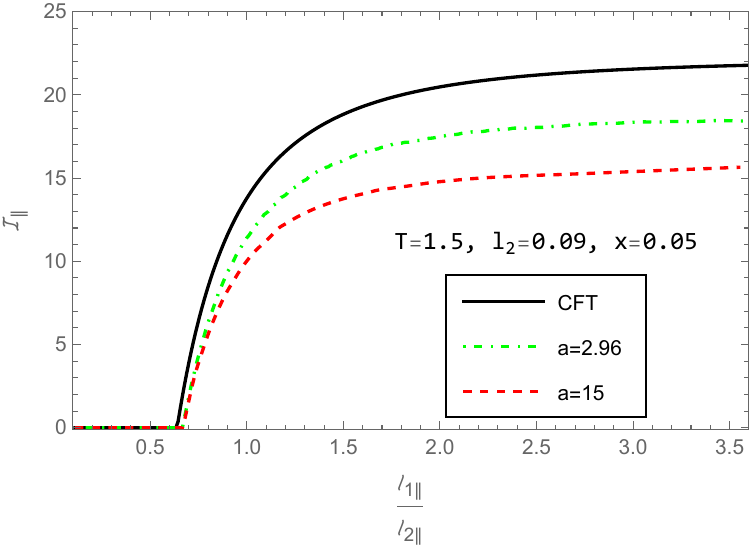}
\hspace{0.5cm}
\includegraphics[width=70mm]{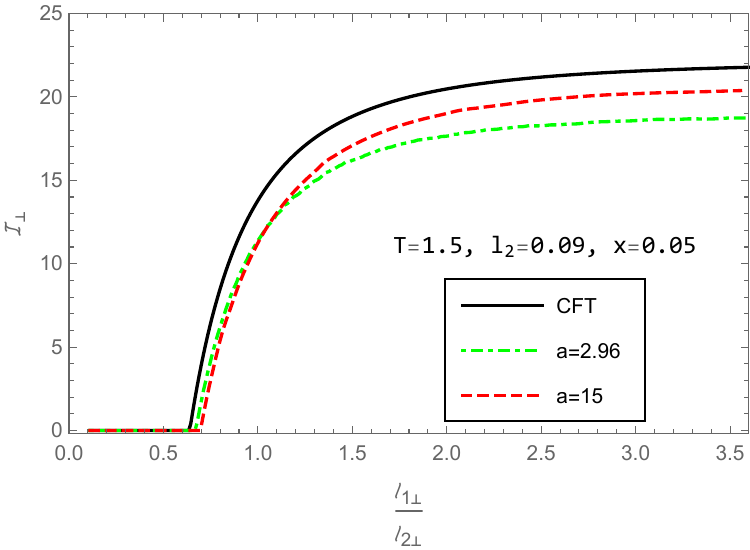}
\caption{Mutual information as a function of $\frac{l_1}{l_2}$ for two strips with unequal lengths at $T=1.5$, for anisotropy parameters $a=2.96$ and $a=15$, with $l_2=0.09$ and $x=0.05$, compared with the isotropic conformal case.}\label{MIunequal}
\end{center}
\end{figure}

Fig.~\ref{MIunequal} shows the mutual information as a function of the ratio $\frac{l_1}{l_2}$ at fixed temperature for different values of the anisotropy parameter. The overall behavior is qualitatively similar to that observed in the equal-length case. Near the phase transition region, where the mutual information approaches zero, increasing the anisotropy parameter suppresses the mutual information, indicating that the increase in energy density is the dominant effect. Away from the critical region, however, the behavior reflects the same competition between energy density and pressure discussed in the previous subsection.

\begin{figure}[h]
\begin{center}
\includegraphics[width=70mm]{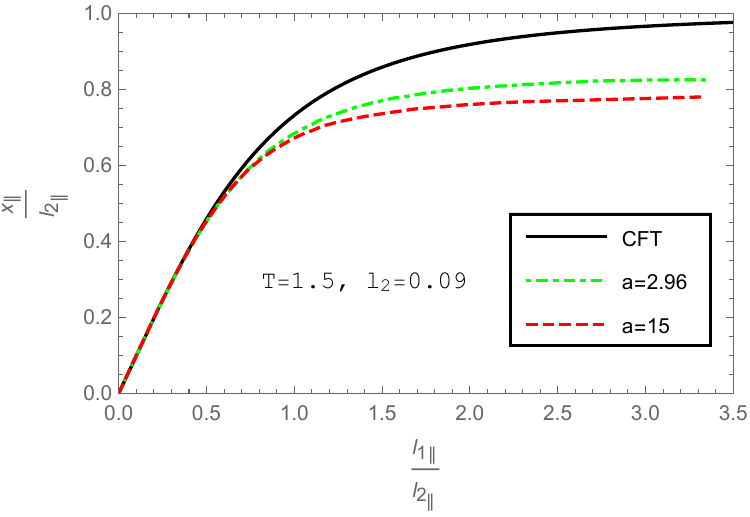}
\hspace{0.5cm}
\includegraphics[width=70mm]{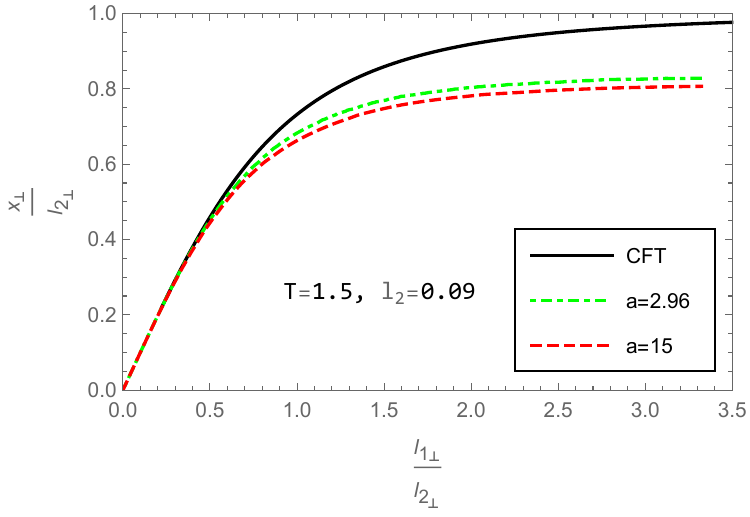}
\caption{Phase diagram for two strips with unequal lengths, plotted as $\frac{x}{l_2}$ versus $\frac{l_1}{l_2}$, at $T=1.5$ for anisotropy parameters $a=2.96$ and $a=15$, with $l_2=0.09$, compared with the isotropic conformal case.} \label{phaseunequal}
\end{center}
\end{figure}

In the perpendicular configuration, the deviation from the isotropic conformal result is gradually reduced and, for sufficiently large values of $\frac{l_1}{l_2}$, larger anisotropy leads to an increase in the mutual information. This behavior reflects the competition between the increasing energy density and the decreasing transverse pressure. In the parallel configuration, by contrast, both the energy density and the longitudinal pressure increase with anisotropy. Consequently, their effects reinforce each other, and the mutual information decreases monotonically with increasing anisotropy.\\
The corresponding phase diagrams are shown in Fig.~\ref{phaseunequal}. As in the equal-length case, increasing the anisotropy parameter reduces the size of the correlated region. Equivalently, the phase boundaries shift toward smaller values of $\frac{x}{l_2}$, implying that the connected phase becomes less favorable. Therefore, stronger anisotropy restricts the parameter range over which the mutual information remains non-zero.

\section{Summary and Outlook}\label{summary}

In this work, we investigated the behavior of holographic mutual information in an anisotropic strongly coupled plasma within the framework of gauge/gravity duality. We considered both parallel and perpendicular strip configurations and analyzed the effects of several parameters, including the anisotropy parameter $a$, temperature $T$, strip lengths, and the separation distance between the subsystems.

For strips with equal lengths, we found that increasing temperature suppresses the mutual information due to enhanced thermal fluctuations and the corresponding increase in energy density. The phase diagrams show that the correlated region becomes smaller as the temperature increases, implying that for a given strip length the corresponding critical separation decreases and the connected phase becomes less favorable.

We also examined the role of anisotropy. In the parallel configuration, increasing anisotropy leads to a monotonic decrease in the mutual information. In contrast, the perpendicular configuration exhibits a richer behavior governed by the competition between the effects of energy density and pressure. Near the phase transition region, the contribution of energy density dominates, leading to a reduction of mutual information. Away from the critical region, however, the pressure effect becomes increasingly important and can enhance the correlations, resulting in larger values of mutual information.

For strips with unequal lengths, we observed qualitatively similar features. In particular, the competition between energy density and pressure gives rise to behavior analogous to that found in the equal-length case. Near the phase transition region, increasing anisotropy suppresses the mutual information, whereas away from the critical region the pressure effect becomes more pronounced and larger anisotropy can lead to stronger correlations. The corresponding phase diagrams indicate that the correlated region decreases as either temperature or anisotropy increases.

Overall, our results demonstrate that temperature and anisotropy play complementary roles in determining the structure of quantum correlations in anisotropic holographic systems. In particular, anisotropy induces nontrivial directional effects and gives rise to a competition between energy density and pressure, which significantly influences the phase structure of mutual information.

As a possible future direction, it would be interesting to extend this analysis to time-dependent backgrounds and investigate the evolution of mutual information during thermalization processes. Studying multipartite information and other nonlocal observables in anisotropic systems may also provide deeper insight into the structure of quantum entanglement in strongly coupled field theories.

\section*{Acknowledgments}
M.R. gratefully acknowledges the hospitality and workspace provided by the Department of Physics at Isfahan University of Technology during the initial stages of this research.
During the preparation and revision of this manuscript, the authors used GPT to assist with English-language editing, consistency checks of the LaTeX notation and references, and the verification and rederivation of selected expressions. All suggestions and outputs were critically reviewed and independently verified by the authors, who take full responsibility for the scientific content and the final form of the manuscript.

\appendix
\section{Different Temperatures at Fixed $\dfrac{a}{T}$}

An interesting question is whether the physical properties of the system are determined solely by the ratio $\dfrac{a}{T}$ or whether they also depend explicitly on the temperature itself. To investigate this issue, we replot the previously discussed quantities for different temperatures while keeping $\dfrac{a}{T}$ fixed.

\begin{figure}[!ht]
\begin{center}
\includegraphics[width=70mm]{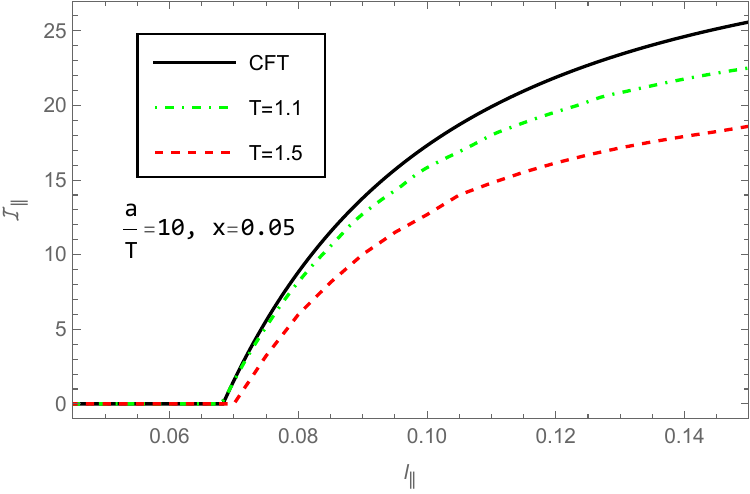}
	\hspace{0.5cm}
\includegraphics[width=70mm]{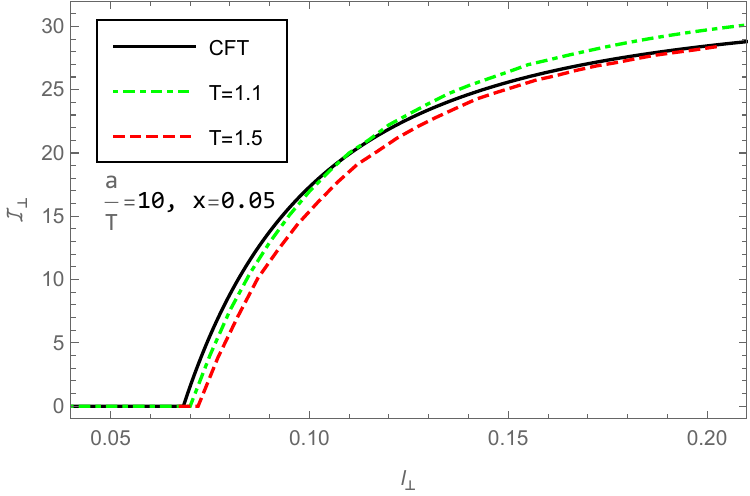}
\caption{Mutual information as a function of the strip length $l$ for fixed $\frac{a}{T}=10$ and separation distance $x=0.05$, at temperatures $T=1.1$ and $T=1.5$, compared with the isotropic conformal case.}\label{MIxaoverT}
\end{center}
\end{figure}

\begin{figure}[!ht]
\begin{center}
\includegraphics[width=70mm]{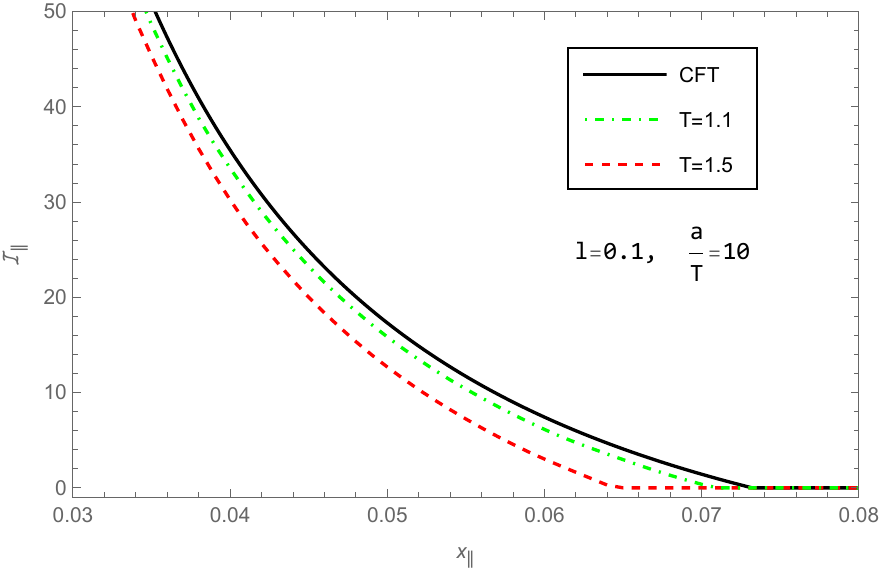}
	\hspace{0.5cm}
\includegraphics[width=70mm]{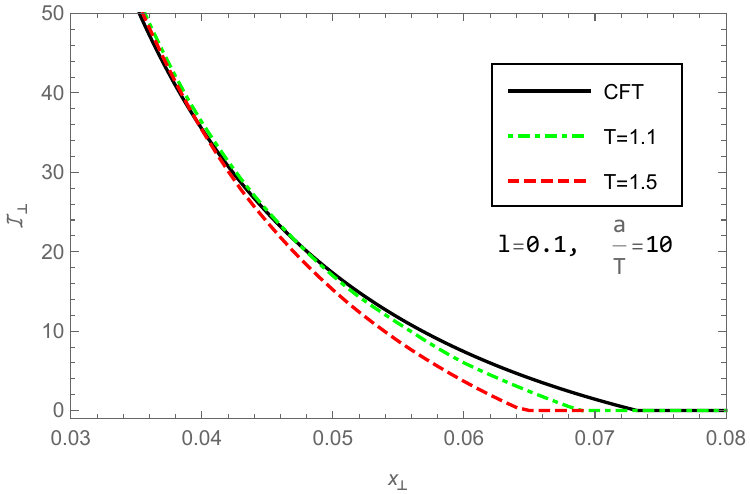}
\caption{Mutual information as a function of the separation distance $x$ for fixed $\frac{a}{T}=10$ and $l=0.1$, at temperatures $T=1.1$ and $T=1.5$, compared with the isotropic conformal case.}\label{MIlaoverT}
\end{center}
\end{figure}

\begin{figure}[!ht]
	\begin{center}
		\includegraphics[width=70mm]{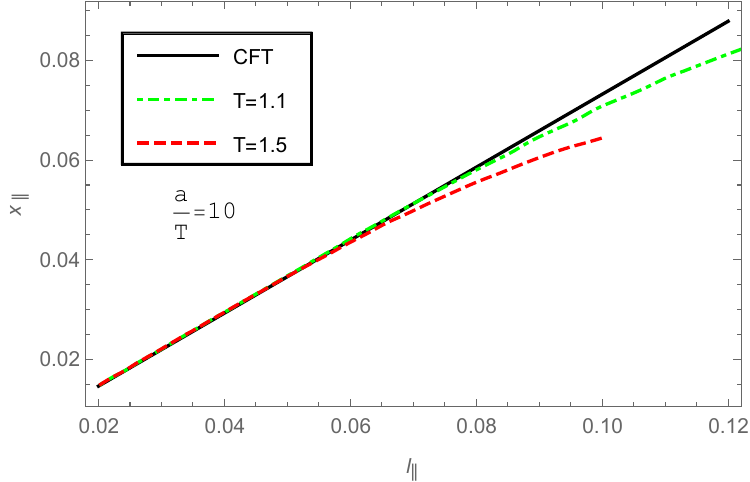}
		\hspace{0.5cm}
		\includegraphics[width=70mm]{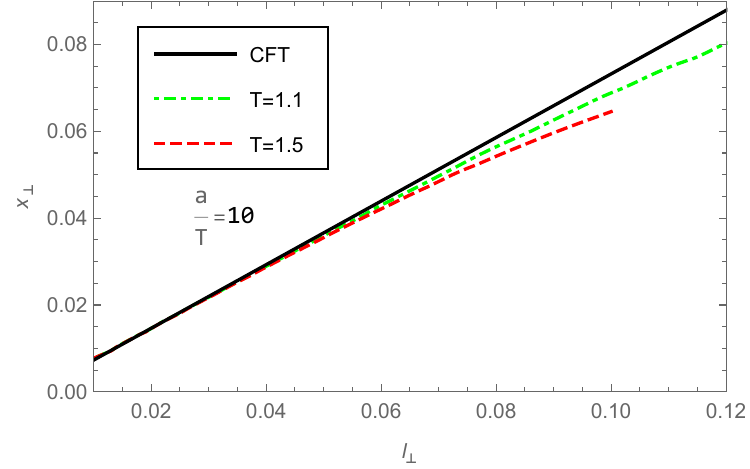}
		\caption{Phase diagram for fixed $\frac{a}{T}=10$, at temperatures $T=1.1$ and $T=1.5$, compared with the isotropic conformal case.}\label{phaseaoverT}
	\end{center}
\end{figure}

As shown in Figs.~\ref{MIxaoverT}, \ref{MIlaoverT}, and \ref{phaseaoverT}, and in agreement with the results of~\cite{Mateos2011}, the mutual information and the phase structure are not determined solely by the ratio $\dfrac{a}{T}$. Even for fixed $\dfrac{a}{T}$, changing the temperature modifies the mutual information and the corresponding phase diagrams, indicating that the physics is not completely characterized by the ratio $\dfrac{a}{T}$ alone.

\section{Low-Temperature Behavior of Mutual Information}

\begin{figure}[h]
\begin{center}
\includegraphics[width=70mm]{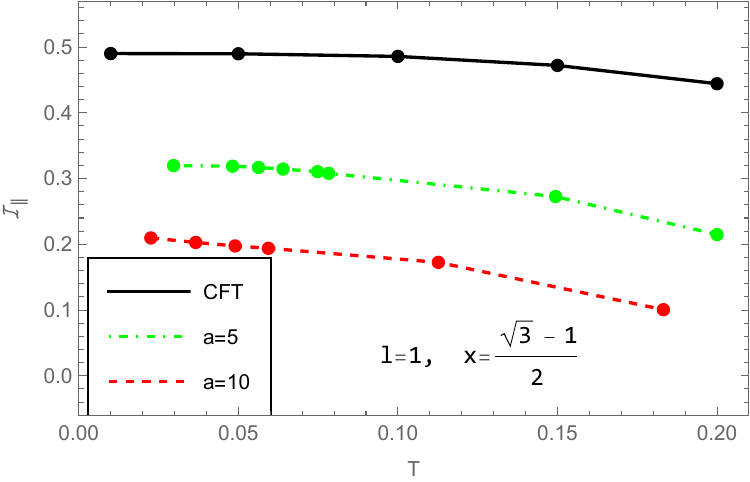}
\hspace{0.5cm}
\includegraphics[width=70mm]{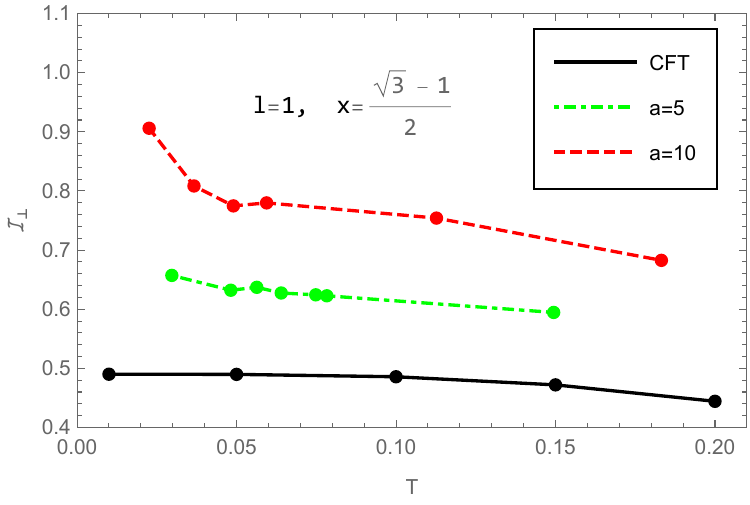}
\caption{Mutual information as a function of temperature at fixed strip length $l=1$ and separation distance $x=\frac{\sqrt{3}-1}{2}$, for anisotropy parameters $a=5$ and $a=10$, compared with the isotropic conformal case.}\label{afix1over2}
\end{center}
\end{figure}

In this appendix, we examine the behavior of the mutual information in the low-temperature regime in order to isolate the effect of anisotropy from thermal effects. In the main text, the temperatures considered are sufficiently high that the overall behavior is largely governed by thermal contributions. By lowering the temperature, the directional effects of anisotropy become more pronounced, allowing the competition between energy density and pressure discussed in the main text to be observed more clearly. Fig.~\ref{afix1over2} shows the mutual information as a function of temperature for fixed strip length $l=1$ and separation distance $x=\frac{\sqrt{3}-1}{2}$, for anisotropic backgrounds with $a=5$ and $a=10$, together with the isotropic conformal field theory.

We observe that the effect of anisotropy depends strongly on the orientation of the strips. In the parallel configuration, the mutual information remains smaller than the isotropic conformal value throughout the low-temperature range, indicating that anisotropy suppresses correlations in this direction. In contrast, the perpendicular configuration exhibits larger mutual information than the isotropic result. At first sight, this behavior may seem at odds with the results presented in the main text. However, it is a natural consequence of the same mechanism discussed there. Increasing the anisotropy parameter raises the energy density, which tends to suppress the mutual information, while simultaneously reducing the transverse pressure, which tends to enhance correlations. At sufficiently low temperatures, where thermal effects are weaker, the pressure contribution can become dominant, leading to larger mutual information in the perpendicular configuration. As the temperature increases, thermal effects gradually dominate, the mutual information decreases, and the behavior continuously approaches that discussed in Figs.~\ref{I(l)fixeda} and \ref{I(x)fixeda}.

These results demonstrate that the effect of anisotropy is not hidden in the main analysis but is partially masked by thermal effects at higher temperatures. When the temperature is sufficiently low, the directional nature of the anisotropic plasma becomes more apparent, making the competition between energy density and transverse pressure clearly visible in the perpendicular configuration.

%\nocite{*}

%\bibliographystyle{IEEEtran}

\bibliographystyle{apsrev4-2}
%\bibliographystyle{plainurlوunsrtnatوIEEEtran}
% سبک رفرنس‌ها (می‌تونی تغییر بدی به چیزی که دوست داری)
%\bibliographystyle{ieeetr}
\bibliography{references2}  % نام فایل .bib بدون پسوند

\end{document}